\documentclass[twocolumn,secnumarabic,amssymb, nobibnotes, aps, prd]{revtex4-2}
\usepackage[T1]{fontenc} 
\usepackage[]{graphicx}
\usepackage{subfigure}
\usepackage{times}

\usepackage{amsmath}
\usepackage{lipsum}
\usepackage{verbatim}

\usepackage{changes}
\usepackage{bm}
\usepackage{ulem}
\usepackage{xcolor}
\usepackage{lineno}
\usepackage{amsmath}

\begin{document}

\title{\LARGE \bf
Strain-Enabled Giant Second-Order Susceptibility in Monolayer WSe$_2$}

\author{Zhizi Guan$^1$, Yunkun Xu$^2$, Junwen Li$^3$, Zhiwei Peng$^2$}

\author{Dangyuan Lei$^{2,}$}

\email{dangylei@cityu.edu.hk}

\author{David J. Srolovitz$^{1,}$}

\email{srol@hku.hk}

\affiliation{\mbox{$^1$Department of Mechanical Engineering, The University of Hong Kong, Pokfulam Road, Hong Kong SAR}\\\mbox{$^2$Department of Materials Science and Engineering, Centre for Functional Photonics, and Hong Kong} Branch of National Precious Metals Material Engineering Research Centre, City University of Hong Kong, Hong Kong SAR \\$^3$DFTWorks LLC, Oakton, VA 22124, USA}

\begin{abstract}
Monolayer WSe$_2$ (ML WSe$_2$) exhibits a high second-harmonic generation (SHG) efficiency under single 1-photon (1-p) or 2-photon (2-p) resonant excitation conditions due to enhanced second-order susceptibility compared with off-resonance excitation states \cite{lin2021narrow,wang2015giant}.   
Here, we propose a novel strain engineering approach to dramatically boost the in-plane second-order nonlinear susceptibility ($\chi_{yyy}$ ) of ML WSe$_2$ by tuning the biaxial strain to shift two K-valley excitons (the A-exciton and a high-lying exciton (HX)) into double resonance. 
We first identify the A-exciton and HX from the 2D Mott-Wannier model for pristine ML WSe$_2$ and calculate the $\chi_{yyy}$ under either 1-p or 2-p resonance excitations, and observe a $\sim$ 39-fold $\chi_{yyy}$ enhancement arising from the 2-p HX resonance state compared with the A-exciton case.
By applying a small uniform biaxial strain (0.16\%), we observe an exciton double resonance state ($E_{\rm{HX}}$ = 2$E_{\rm{A}}$, $E_{\rm{HX}}$ and $E_{\rm{A}}$ are the exciton absorption energies), which yields up to an additional 52-fold enhancement in $\chi_{yyy}$ compared to the 2-p HX resonance state, indicating an overall $\sim$ 2000-fold enhancement compared to the single 2-p A-exciton resonance state reported in Ref \cite{wang2015giant}.
Further exploration of the strain-engineered exciton states (with biaxial strain around 0.16\%) reveals that double resonance also occurs at other wavevectors near the K valley, leading to other enhancement states in $\chi_{yyy}$, confirming that strain engineering is an effective approach for enhancing $\chi_{yyy}$. 
Our findings suggest new avenues for strain engineering the optical properties of 2D materials for novel nonlinear optoelectronic applications.


\end{abstract}

\maketitle




\begin{figure}[htb]
  \centering
  \subfigure{\includegraphics[scale=0.42]{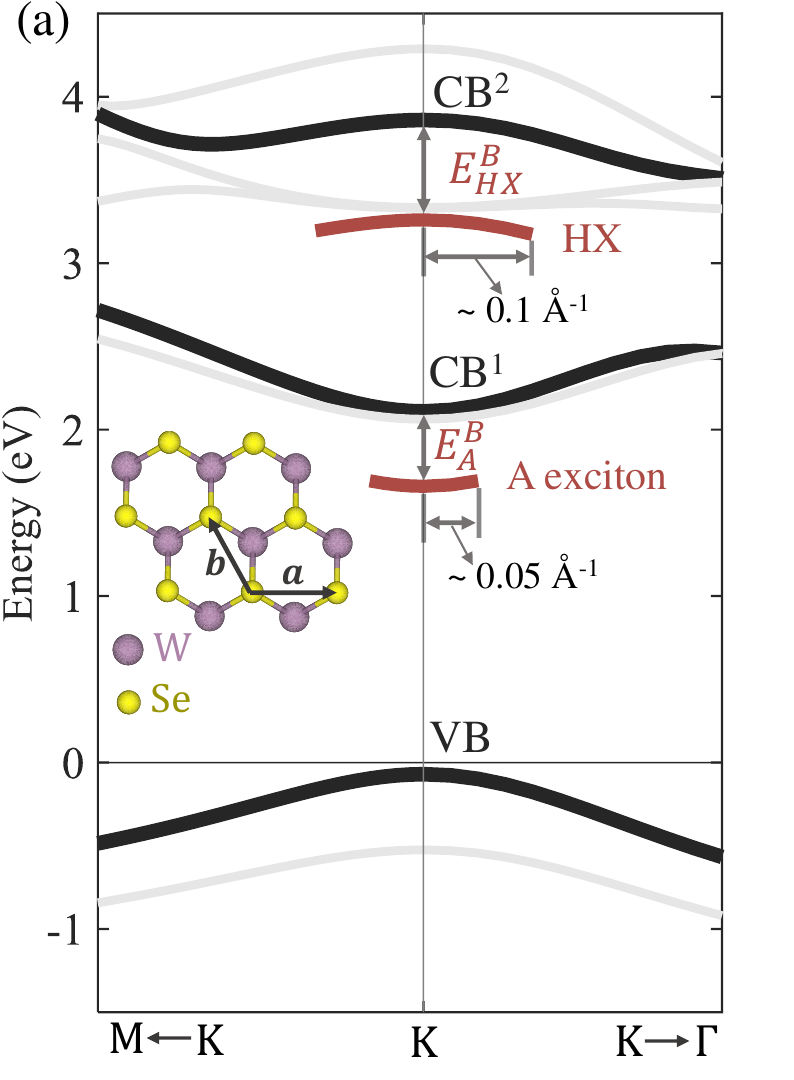}}
  \hfill
  \subfigure{\includegraphics[scale=0.365]{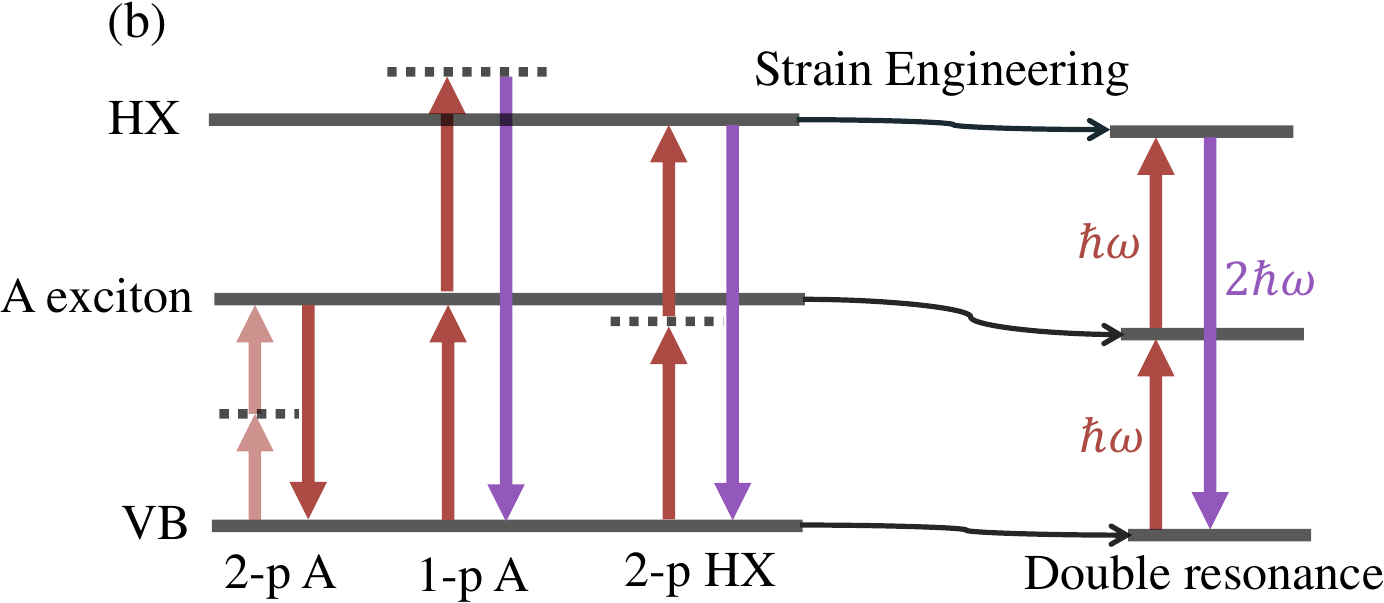}}
  \caption{(a) Calculated band structure (near the K valley) of ML WSe$_2$, with its crystal structure illustrated in the inset. Three spin-split bands (VB, CB$^1$, CB$^2$) associated with the A and HX exciton states are highlighted in bold. Two exciton states are  labeled based on their envelope functions from reference \cite{lin2021narrow}, with their binding energies ($E_{\rm{A}}^B$, $E_{\rm{HX}}^B$) according to our 2D Mott-Wannier model calculations. 
(b) Schematics of 1-photon (1-p), 2-photon (2-p), and double resonance states in ML WSe$_2$ based on A-exciton and HX. The dashed and solid lines depict the SHG virtual states and real exciton states, respectively. Single 1-p or 2-p resonance requires that the incident energy or twice the incident energy meet the excitonic energy, while double resonance occurs when both energy matching requirements are met, which could be realized by strain engineering.}
  \label{fig:schmetic}
\end{figure}

Second harmonic generation (SHG) is a process in which incident photons interact with a nonlinear, noncentrosymmetric crystal to generate new photons with twice the initial frequency \cite{franken1961generation, zhang2020second}. 
SHG finds significant applications in monolayer transition metal dichalcogenides (TMDCs) partially due to their atomic thickness induced relaxation of phase matching constraints between the incident and second-harmonic waves \cite{boyd2020nonlinear,khan2022optical,xie2024nonlinear,ciattoni2018phase}.
Physically, the SHG intensity is proportional to the square of the material’s second-order nonlinear optical susceptibility ($\bm{\chi^{(2)}}$). 
An important strategy to enhance $\bm{\chi^{(2)}}$ is through one-photon (1-p) or two-photon (2-p) resonant excitation, which entails the alignment of SHG virtual states with the eigenenergy levels of a material \cite{lin2021narrow,Yao2020,song2018second,le2017impact,wang2015giant,shree2021interlayer,dai2020electrical,seyler2015electrical}.
In monolayer TMDCs, the resonance state can be achieved by matching the incident 1-p or 2-p energy with an exciton energy.
Based on this mechanism, Lin et al.  \cite{lin2021narrow} reported a narrow and intense SHG peak from a 716 nm continuous-wave (CW) laser resonantly driving the A-exciton in ML WSe$_2$. 
The observed intense SHG signal from a relatively low pump irradiance, which is 2-3 orders of magnitude lower than pulsed laser commonly used in nonlinear optical characterization of monolayer materials, provides strong evidence of the 1-p A-exciton resonance induced a large second-order susceptibility \cite{Yao2020}. 
Wang et al.  \cite{wang2015giant} reported a 2-p resonance state in ML WSe$_2$ by targeting 2-photon excitation energy at the A-exciton, leading to an up to  3 orders of magnitude enhancement in SHG efficiency (compared with the off-resonance state).

%

To further enhance the second-order susceptibility, we propose to combine 1-p and 2-p resonance processes at a single wavevector; i.e. "double resonance"  (Fig.~\ref{fig:schmetic}(b)). 
Double resonance is a promising  approach to inducing a giant enhancement in the second-order susceptibility compared to single 1-p or 2-p resonance state \cite{duan1999first,sipe2000second,cabellos2009effects,trolle2014theory,brun2015intense,biswas2023double}. 
To our knowledge, the double resonance scenario has never been observed in real experiments due to its restrictive band configuration criteria: three bands must form two continually nested states (in reciprocal space); this is an extension of the two-band nesting concept in first-order linear absorption processes \cite{bassani1976electronic,carvalho2013band,mennel2020band}.
Fortunately, a recent experiment identified a novel bright exciton state with an in-plane dipole orientation and negative effective mass, termed "high-lying exciton" (HX), at the K valley \cite{lin2021narrow}. Remarkably, the energy of the HX is close to twice that of the A-exciton in ML WSe$_2$, rendering it an ideal candidate for achieving double resonance combined with the A-exciton. 

In the above context, we investigate the potential of the high-lying exciton state for boosting the in-plane second-order susceptibility ($\chi_{yyy}$, a tensor component of $\bm{\chi^{(2)}}$) of ML WSe$_2$ through double resonance with A-exciton state. 
In this letter,  we explore the application of biaxial strain ($\eta$) engineering to tune the band energies around the K valley and two exciton states in order to realize the double resonance condition for $\chi_{yyy}$ giant enhancement. 
First, we identify the exciton binding energy and the absorption peaks ($E_{\rm{A}}$, $E_{\rm{HX}}$)  within a 2D Mott-Wannier model and first-principle calculations. 
We then compare the single 1-p and 2-p resonance state enhanced $\chi_{yyy}$ and find that the strongest on-resonance $\chi_{yyy}$ comes from the 2-p resonance state at HX in unstrained ML WSe$_2$, manifested by a 39-fold enhancement compared to the 2-p resonance state at A-exciton. 
By applying a biaxial tensile strain  ($\eta$ = 0.16\%) to satisfy the excitonic double resonance condition ($E_{\rm{HX}}$ = 2$E_{\rm{A}}$), we observe an additional 52-fold enhancement in $\chi_{yyy}$ compared to the single 2-p HX resonance state, leading to an overall three orders of magnitude enhancement in $\chi_{yyy}$.
By exploring different strain states to achieve double resonance at other wavevectors near the K valley (0.16\% $\leq\eta$ $\leq$ 0.67\%), the double resonance effect also enhances $\chi_{yyy}$ compared to the single 1-p or 2-p resonance state. 
Our findings demonstrate the potential of strain engineering in 2D materials for optimizing their nonlinear optical properties, holding considerable potential in different nonlinear optoelectronic applications. 


We obtain the optimized crystal structures and self-consistent wavefunctions using density functional theory (DFT) within the projected augmented wave (PAW) formalism implemented in \textit{VASP} \cite{kresse1996efficient,perdew1996generalized,blochl1994projector}. 
We compare the Perdew-Burke-Ernzerhof (PBE) and Heyd–Scuseria–Ernzerhof (HSE) hybrid functionals for describing electron exchange-correlation, with previously reported self-consistent GW results  \cite{heyd2003hybrid,heyd2004efficient}. 
Our main results are based on the PBE scheme; the PBE and HSE band shapes exhibit negligible difference near the K valley (see Fig.~S1), while the choice of PBE effectively reduces the computational demands compared with HSE and GW-based calculations. 
Neither PBE nor HSE band structures compare well with GW results using only a single-step scissor correction at the band gap. 
Therefore, we apply different corrections for different conduction band valleys to ensure that the corrected bands align well with the previously reported GW results \cite{lin2021narrow} (scissor correction details and comparisons among exchange-correlation schemes are provided in Sections I and II of the SM).

We determine the second-order susceptibility tensor ($\bm{\chi^{(2)}}$) within the independent-particle approximation (IPA) while with adequate correction, which well describes our target excitons (A-exciton, HX) for discussion. 
In addition to the PAW-PBE, we also describe electron and hole wavefunctions using the maximally localized Wannier function (MLWF) basis \cite{mostofi2008wannier90}. 
Comparisons and computation details may be found in Section IV of the SM. 
Wannier interpolation is utilized to efficiently compute  energy and momentum matrix elements on an arbitrary $\bm{k}$ grid in reciprocal space. 
It is important to note that achieving double resonance  leads to numerical difficulties (divergences) in evaluating the second-order susceptibility  using regular Monkhorst-Pack sampling. 
Hence, we employ a  method in which the target Brillouin zone is divided into triangles across which  the energy and momentum matrix elements are linear within each triangle \cite{brun2015intense,wiesenekker1988analytic} to achieve faster convergence and improved accuracy  of our second-order susceptibility results (see Section V, Fig.~S5, S6 of the SM).

\begin{figure}[htb]
  \centering
  \subfigure{\includegraphics[scale=0.275]{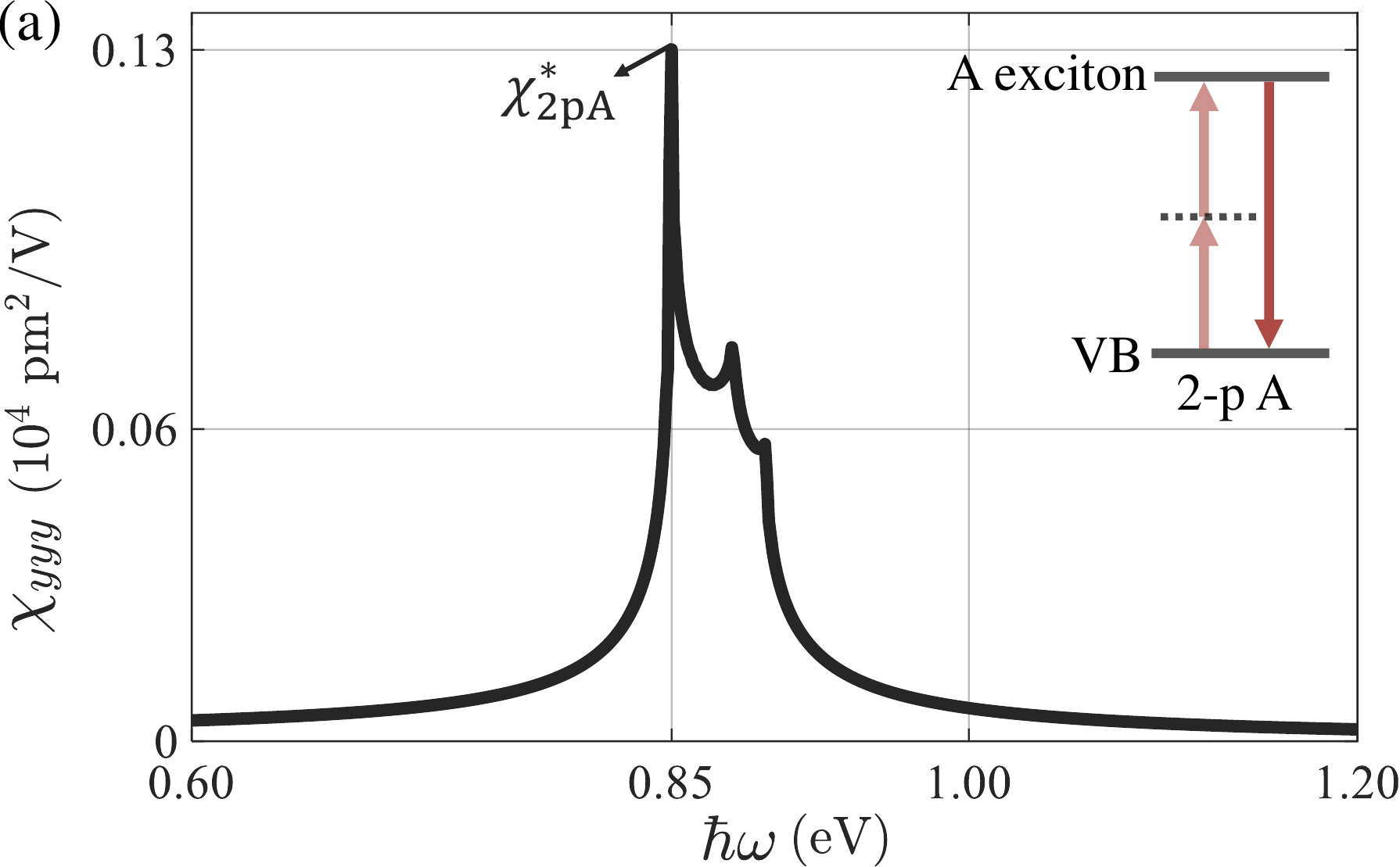}}
  \hspace{0.1in}
  \subfigure{\includegraphics[scale=0.28]{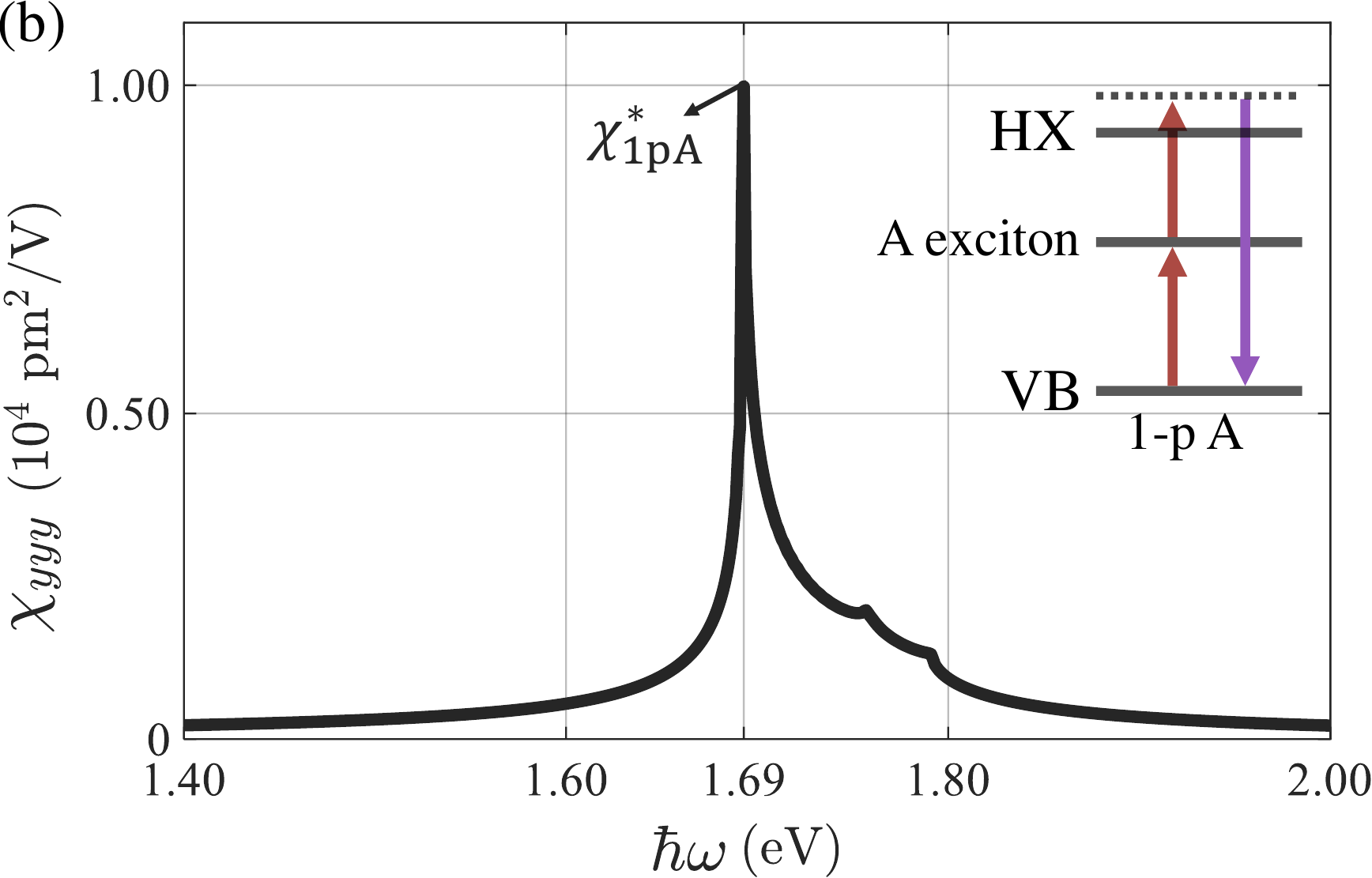}}
  \hspace{0.1in}
  \subfigure{\includegraphics[scale=0.28]{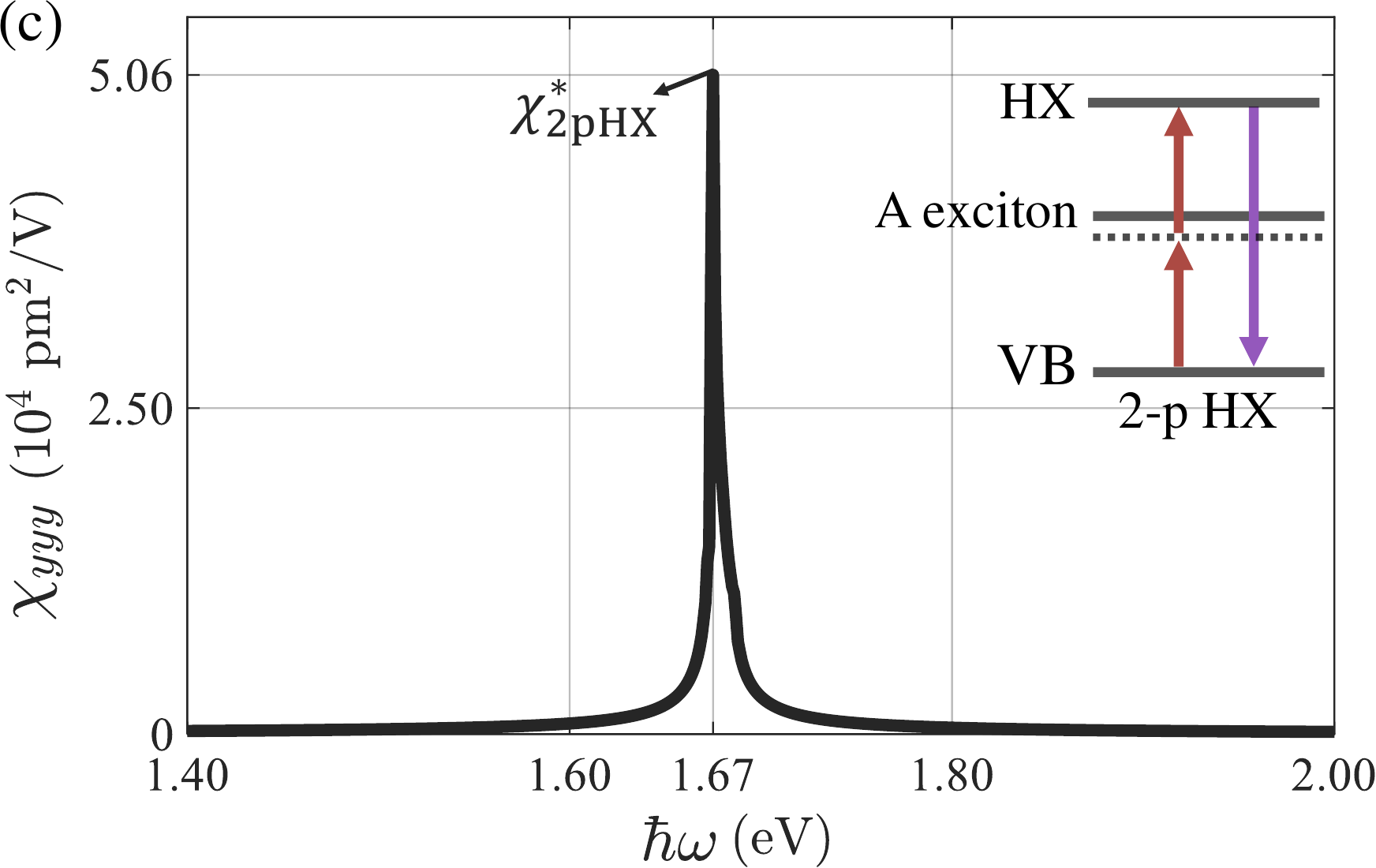}}
  \caption{Calculated in-plane second-order nonlinear susceptibility $\chi_{yyy}$ (absolute value) versus incident photon energy $\hbar\omega$ at the single 1-p (b) or 2-p resonance state (a \& c). We label the peak value at each on-resonance incident energy as (a) $\chi_{\rm{2pA}}^\ast$, (b) $\chi_{\rm{1pA}}^\ast$, and (c) $\chi_{\rm{2pHX}}^\ast$ to compare the enhancement among different states.}
  \label{fig:exciton}
\end{figure}


Our first step is to identify the A-exciton and HX based on the scissor-corrected band structure and the Mott-Wannier model. 
We use the experimentally-defined lattice constant (3.282 \AA) in our calculation model for direct comparison with previously reported experimental results. 
Fig.~\ref{fig:schmetic}(a) illustrates the atomic structure and band structure of ML WSe$_2$, exhibiting a direct band gap at the K valley of 2.18 eV. We then determine the exciton binding energy using the Mott-Wannier model,   regarding the electron-hole pair as a  "pseudo-hydrogen atom" \cite{wannier1937structure}. 
The 3D Mott-Wannier exciton model must be modified to account for  the  non-local and wavevector-dependent screening  in 2D materials. 
Hence, we employ the 2D Mott-Wannier model \cite{olsen2016simple} to calculate the binding energy as $E^{B}=(8 \mu)/(1+\sqrt{1+32 \pi \alpha \mu / 3})^2$, where $\mu$ represents the reduced effective mass of the electron-hole pair forming the exciton ($\mu^{-1}=m_e^{-1}+m_h^{-1}$). 
$\mu$  can be  fit from the band structure as ${\hbar^2}{(d^2 E(\bm{k}) / d \bm{k}^2)}^{-1}$, where $\hbar$ is the reduced Planck constant and $E(\bm{k})$ is the energy band dispersion in  momentum space. 
$\alpha$ denotes the internal polarizability, which is linearly fit to the wavevector-dependent dielectric constant. 

In ML WSe$_2$, the A-exciton and HX can be depicted as an electron and hole in three bands forming two pairs in Fig.~\ref{fig:schmetic}(a) (A: VB, CB$^1$; HX: VB, CB$^2$). 
Based on the  PBE functional results, the A  and HX  exciton binding energies are 0.48 eV and 0.58 eV in ML WSe$_2$, respectively, in good agreement with reported BSE results (A: 0.45 eV, HX: 0.60 eV). Therefore, the absorption energies of A ($E_{\rm{A}}$) and HX ($E_{\rm{HX}}$) in the unstrained ML WSe$_2$ are at 1.70 eV and 3.35 eV, respectively, by subtracting each binding energy from the GW gap at the K valley. By calculating the oscillator strength (${|<\psi_v|\hat{p}| \psi_c>|^2}/{(E_c-E_v)}$, v$\in$\{VB\}  c$\in$\{CB$^1$, CB$^2$\}) at the K valley for the two excitons in Fig.~S9 (c), we found that the oscillator strength ratio also correspond with previous BSE results (i.e., the intensity ratio HX: A $\approx$ 1:18), but at much lower computational cost \cite{lin2021narrow} (Section III of the SM). In summary, our PBE-based description of two excitons, with appropriate corrections, enables a good representation of the A-exciton and HX for subsequent strain engineering.

Based on the identification of two excitons, we explore how single 1-p and 2-p resonance from the A-exciton and HX enhances $\chi_{yyy}$. Fig.~\ref{fig:exciton}(a)-(c) separately shows three single resonance effects near the resonance incident energy, and we label the highest on-resonance $\chi_{yyy}$ peak as $\chi^{\ast}$ to observe the maximum enhancement fold (we isolated the 1-p and 2-p states in the $\chi_{yyy}$ calculation - see Section V, SM and Fig.~S7). Due to the relatively small value of the 1-p HX peak and  its uV emission energy, we focus  on the first three resonance states. Fig.~\ref{fig:exciton}(a) shows as an example describing the 2-p A resonance. 
We clearly observe that $\chi_{yyy}$  is 30 to 40-fold that of the off-resonance state susceptibility at $\hbar\omega$ = 0.85 eV  (i.e., $\chi_{2pA}^{\ast}$),   
in agreement with the previous report  3 orders of magnitude enhancement of the SHG efficiency from 2-p A resonance \cite{wang2015giant}. 
Among the three enhancement states, the strongest  comes from 2-p HX resonance; this is likely because the intermediate virtual state energy is nearly aligned with the A-exciton energy, leading to a state that is  closest to double resonance amongst the three. Hence, we use the 2-p HX resonance result to benchmark the double resonance enhancement effect in the next step.

\begin{figure}[htb]
  \centering
  \subfigure{\includegraphics[scale=0.34]{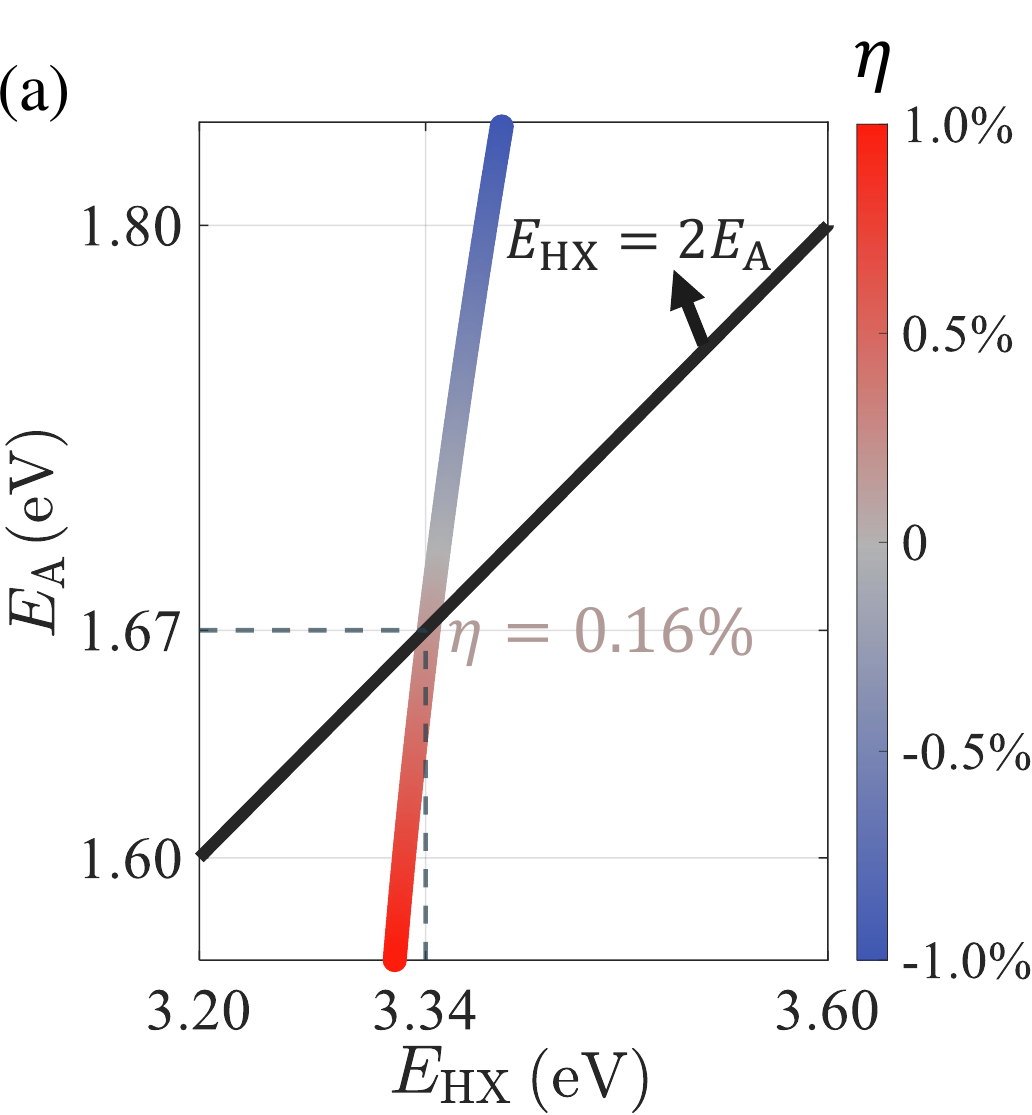}}
  \hspace{0.02in}
  \subfigure{\includegraphics[scale=0.28]{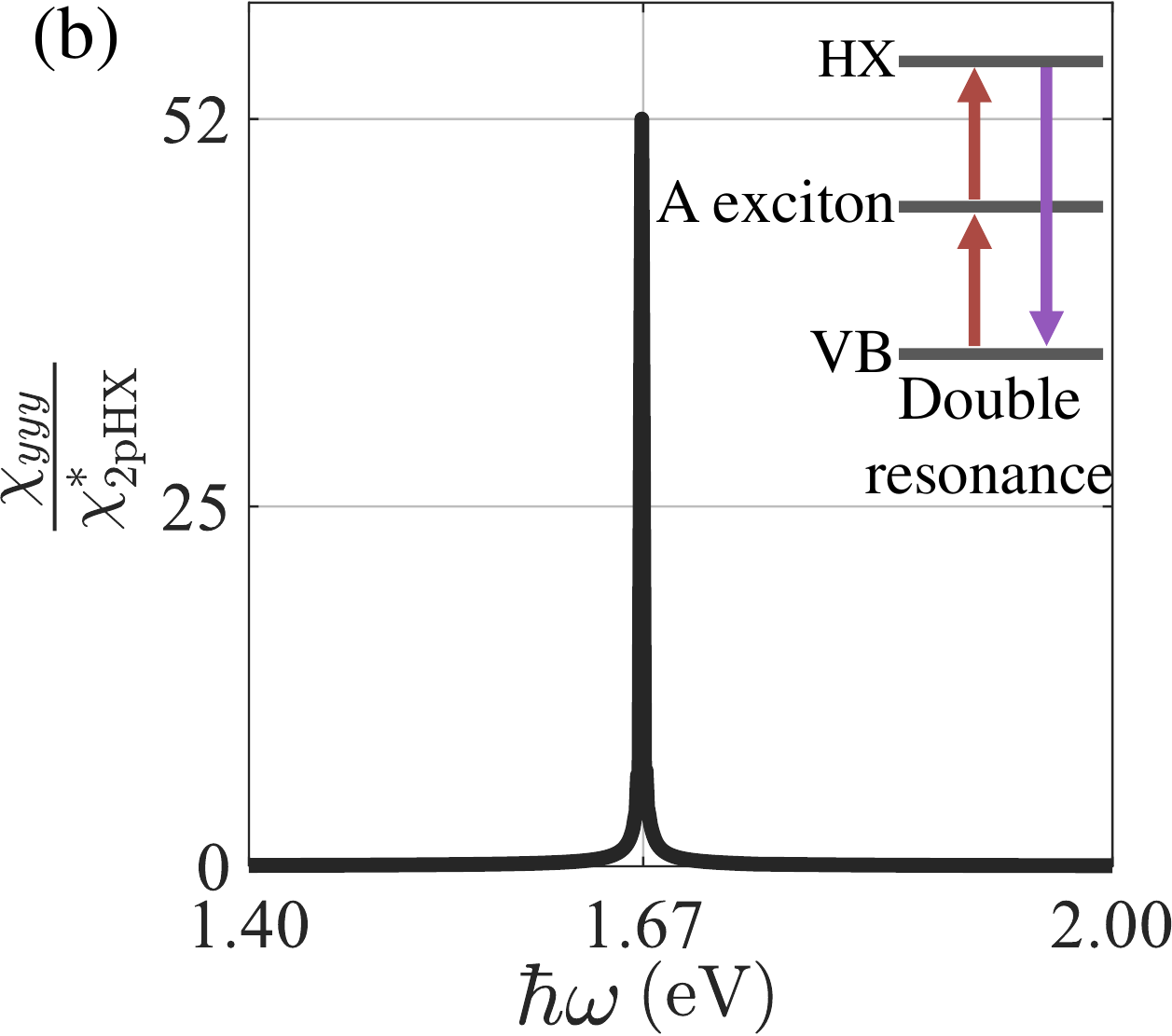}}
  \hspace{0.02in}
  \caption{(a) Calculated excitonic absorption energies $E_{\rm{HX}}$ and $E_{\rm{A}}$ as a function of biaxial strain, $\eta$ (colored line; the color bar on the right represents the biaxial strain, $\eta$). 
The dark line illustrates the double resonance condition, $E_{\rm{HX}} = 2E_{\rm{A}}$. The crossing of the dark and colored line indicates the occurrence of double resonance at $\eta = 0.16$\%. 
(b) The in-plane second-order susceptibility, $\chi_{yyy}$, as a function of incident photon energy at  double resonance  ($\eta = 0.16$\%). An $\sim50$-fold enhancement appears at the double resonance on-resonance incident energy $\sim1.67$ eV compared to the 2-p HX on-resonance state.}
  \label{fig:2DEnergy}
\end{figure}

We then proposed to tune the energies of the A-exciton and HX into excitonic double resonance;  i.e., to ensure that  $E_{\rm{HX}}$ = 2$E_{\rm{A}}$ where $E_{\rm{HX}}$ and $E_{\rm{A}}$ are the absorbance energies for these two excitons. Without external tuning, $E_{\rm{HX}}$ $<$ 2$E_{\rm{A}}$ in ML WSe$_2$. We tune these energy levels via strain engineering (via easy to implement biaxial  strain)  to shift the bands relative to the Fermi level and the excitonic levels relative to each other~\cite{guan2022direct}. Fig.~\ref{fig:2DEnergy}(a) shows how $E_{\rm{A}}$ and $E_{\rm{HX}}$ respond to strain ($\eta$, -1\% $\le$ $\eta$ $\le$ 1\%) in ML WSe$_2$. They both redshift with increasing $\eta$, consistent with  experimental reports \cite{aslan2018strain,peng2020strain}. To interpret the variation trend, we examine the effect of strain $\eta$ on the VB, CB$^1$ and CB$^2$ band energies ($E^o_{v}$, $E^o_{c1}$, $E^o_{c2}$) and the exciton binding energies  ($E^B_{\rm{A}}$, $E^B_{\rm{HX}}$) at the K valley  (Fig.~S9). The relative shift rate  between the CB$^1$ and VB is almost four times larger than that between CB$^2$ and VB due to different bonding nature in different bands (Fig.~S10). And the A-exciton and HX binding energy $E^B$  decreases with increasing $\eta$. This may be attributed to more spatially separated electron-hole pairs, with weaker Coulomb interactions, with increasing $\eta$.


Fig.~\ref{fig:2DEnergy}(a)  illustrates the relationship between $E_{\rm{A}}(\eta)$ and $E_{\rm{HX}}(\eta)$ as well as the excitonic double resonance condition ($E_{\rm{HX}}$ = 2$E_{\rm{A}}$).
This condition is satisfied at $\eta = 0.16$\%, where $E_{\rm{A}}$ = 1.67 eV and $E_{\rm{HX}}$ = 3.34 eV. 
 By calculating the relative enhancement  ${\chi_{yyy}}/{\chi^\ast_{\rm{2pHX}}}$ under the double resonance condition, we see (Fig.~\ref{fig:2DEnergy}(b))  that this strain leads to an up to $\sim50$-fold enhancement in $\chi_{yyy}$ compared with the single 2-p HX resonance result in unstrained ML WSe$_2$. 
This $\sim50$-fold enhancement in $\chi_{yyy}$ indicates an overall $\sim$ 2000-fold enhancement compared to the single 2-p A-exciton resonance state reported in Ref \cite{wang2015giant}, making ML WSe$_2$ considerably potent for many nonlinear optoelectronic applications. 


\begin{figure}[htb]
  \centering
  \subfigure{\includegraphics[scale=0.4]{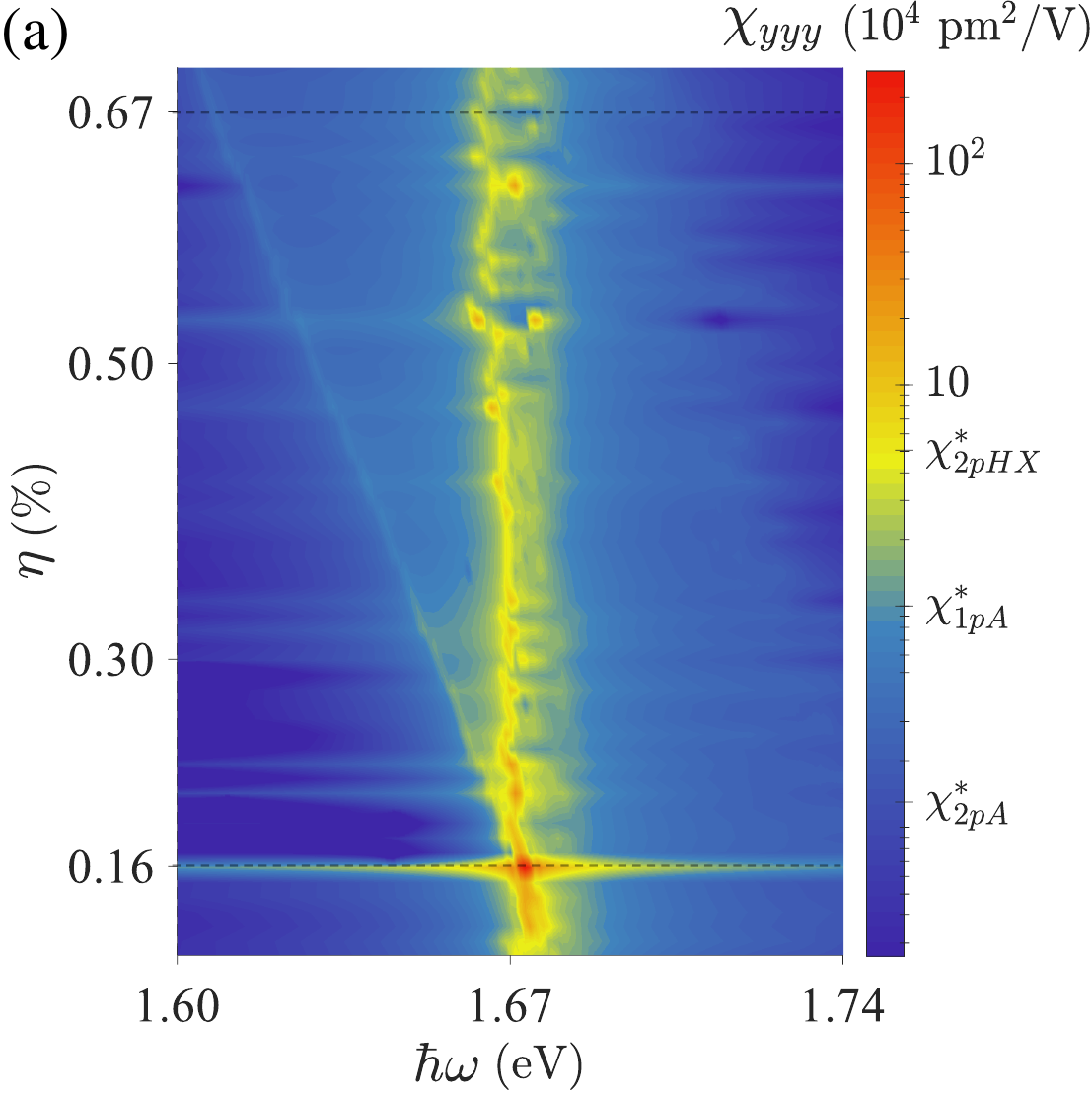}}
  \hspace{0.05in}
  \subfigure{\includegraphics[scale=0.33]{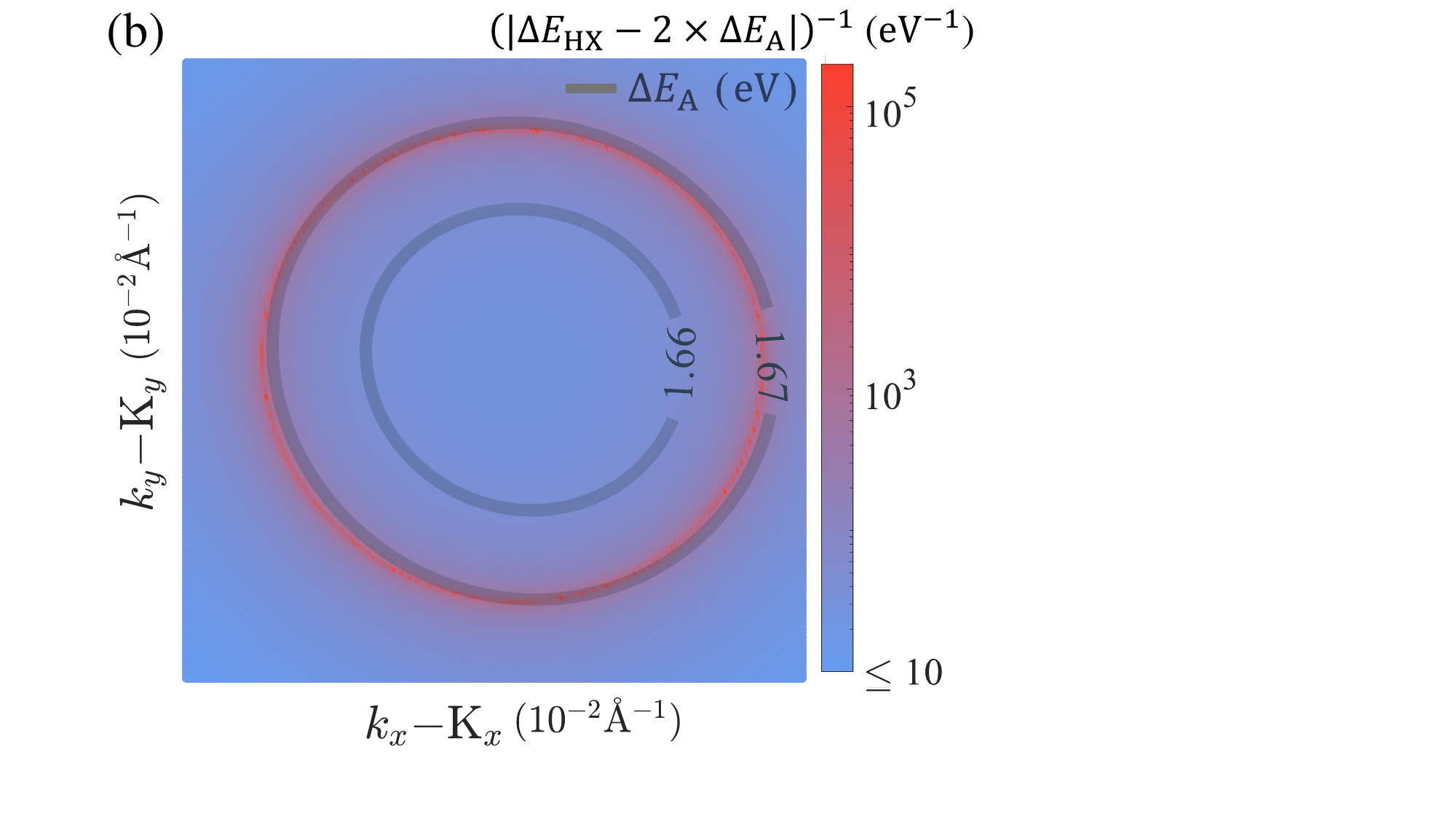}}
  \caption{(a) The absolute value of $\chi_{yyy}$ verses the biaxial strain $\eta$ in the range where double resonance can occur near the K valley (in the DFT predicted  range,   0.16\% $\leq\eta$ $\leq$ 0.67\%). 
Three on-resonance enhancements from single 1-p and 2-p resonance are labeled in the colorbar for easy comparison. 
(b) The double resonance condition $|\Delta E_{HX} - 2\Delta E_{A}|^{-1}$ for one biaxial strain $\eta = $ 0.32\%  ($\Delta E_A$ = $E^o_{c1}$-$E^o_{v}$-$E^B_{\rm{A}}$; $\Delta E_{HX}$= $E^o_{c2}$-$E^o_{v}$-$E^B_{\rm{HX}}$). 
The overlap between the $\Delta E_{A} = 1.67$ eV (gray) contour and the region where $|\Delta E_{HX} - 2\Delta E_{A}|^{-1}$ is large  shows the double resonance effect beyond the K valley.}
  \label{fig:SHG}
\end{figure}


The $50$-fold enhancement in $\chi_{yyy}$ (based on  single 2-p HX resonance) is an important tuning outcome. 
More enhancement states can be anticipated through strain-tuning the exciton energy over the entire exciton band, as other double resonance states are possible in addition to the K valley. 
Hence we first examine other strain-induced $\chi_{yyy}$ enhancement - see Fig.~\ref{fig:SHG}(a). 
As expected, the strongest enhancement  appears near $\hbar\omega$ = 1.67~eV; i.e., the  K-valley double resonance energy. 
Besides this, other magnitudes of biaxial strain lead to an enhanced $\chi_{yyy}$ relative to the single on resonance $\chi_{yyy}$ (i.e.,  $\chi_{\rm{2pHX}}^{\ast}$, $\chi_{\rm{1pA}}^{\ast}$, and $\chi_{\rm{2pA}}^{\ast}$) in the unstrained monolayer. 
Fig.~\ref{fig:SHG}(a) demonstrates that strain tuning is an effective approach for enhancing the second-order susceptibility  compared to  previously reported single 1-p and 2-p resonance methodologies.  

To explain the strain tuning effect outside of the K valley, we explore the exciton energies from the band energy and corresponding binding energy in the reciprocal space, i.e. $\Delta E_{\rm{A}}$ = $E^o_{c1}$-$E^o_{v}$-$E^B_{\rm{A}}$, $\Delta E_{\rm{HX}}$= $E^o_{c2}$-$E^o_{v}$-$E^B_{\rm{HX}}$. 
We quantify the A-exciton energy $\Delta E_{\rm{A}}(\bm{k}, \eta)$ by  fitting the 2D $E$-$\bm{k}$ diagram to a circle as $\Delta E_A(\bm{k},\eta)=\Delta E_A(\rm{K}, \eta)+({\hbar^2}/{2m_{c1}(\eta)}-{\hbar^2}/{2m_v(\eta)})(k_x^2+k_y^2)$, where $m_{c1}$ and $m_v$ are the CB$^1$ and VB effective masses in the K valley (Fig.~S11, S12).  $\Delta E_{\rm{\rm{HX}}}$ exhibits a much smaller variation ratio (with larger anisotropy) than $\Delta E_{\rm{A}}$ (Fig. S13). Hence, we approximate $\Delta E_{\rm{HX}}$($\bm{k}$, $\eta$) = $\Delta E_{\rm{HX}}$(K, $\eta$), where $\Delta E_{\rm{HX}}$(K, $\eta$) is the HX band energy difference in the K valley.


To achieve double resonance,  $\Delta E_{\rm{HX}}(\bm{k}, \eta) = 2\Delta E_A(\bm{k}, \eta)$ must have a solution $\eta(\bm{k})$ in $\sqrt{k_x^2 + k_y^2} \le 5 \times 10^{-2}$ \AA$^{-1}$ (where both excitonic envelope functions are greater than 0). 
This occurs for $0.16\% \le \eta \le 0.67\%$ (dashed lines in Fig.~\ref{fig:SHG}(a)). At $\eta = 0.16\%$, only the K-valley wavevector satisfies the double resonance condition, leading to the strongest enhancement in $\chi_{yyy}$. 
As $\eta$ increases from 0.16\%, the double resonance condition is met at other wavevectors, forming a circle of wavevectors satisfying double resonance. 
At $\eta = 0.67\%$, the double resonance wavevector circle reaches the edges of the A-exciton range. 
We illustrate this by plotting the energy of $|\Delta E_{\rm{HX}}-2\Delta E_{\rm{A}}|^{-1}$ and  the $\Delta E_A$ contour versus $\bm{k}$ on the same figure, in Fig.~\ref{fig:SHG}(b) (for $\eta$ = 0.32\%). 
We observe a circle of overlapping states around the K valley; $|\Delta E_{\rm{HX}}-2\Delta E_{\rm{A}}|^{-1}$ is a maximum (red) and overlaps the  $\Delta E_{\rm{A}}$ = 1.67 eV contour. 
This  proves the existence of double resonance near the K valley, corresponding with the $\chi_{yyy}$ result in Fig.~\ref{fig:SHG}(a) and Fig.~S8 for the  2-fold enhancement of  $\chi_{yyy}$ peak intensity  (compared with $\chi_{\rm{2pHX}}^{\ast}$  of the unstrained monolayer) for an incident photon energy of 1.67 eV.


Our first-principle calculations  indicate that  a small strain can produce a very large enhancement of the second-order susceptibility and  the corresponding SHG output, reminiscent of previous experiments for different monolayer materials. 
We find that different experimental measurements of second-order susceptibility in ML TMDCs yield very different results (up to two-orders of magnitude) \cite{pike2021second,ribeiro2015second,rosa2018characterization}.
We speculate that this difference may arise from the strain-induced resonance effect when ML TMDCs is placed on a substrate for SHG measurements. 
Rahman et al. \cite{rahman2022extraordinary} estimated that a 0.35\% strain was presented in a wrinkled monolayer of CuInP$_2$S$_6$ (CIPS) and showed that this led to an $160$-fold increase in the SHG intensity (which was sensitive to incident photon energy). 
Their experimental observation of a giant SHG enhancement  may indeed be the result of the excitonic resonance mechanism suggested here.
This provides further impetus for further experiments as a function of  strain (and possibly strain gradients) to engineer the second-order susceptibility.




In conclusion, we explored the potential of exploiting high-lying exciton double resonance (with the A-exciton) to induce a giant enhancement of utmost three orders of magnitude in second-order susceptibility ($\chi_{yyy}$) of ML WSe$_2$, compared with the single 1-p or 2-p resonance state. 
We further demonstrated that  biaxial strain is an effective approach to tune the band energies and  excitonic states into double resonance. 
We identify two excitonic absorbance peaks ($E_{\rm{A}}$, $E_{\rm{HX}}$) and their binding energies ($E^B$) within the 2D Mott-Wannier model. 
Our results exhibit good agreement with those from previous  Bethe-Salpeter equation BSE results (with much lower computational demands) and experiments. 
Comparing the 1-p and 2-p resonance state enhanced $\chi_{yyy}$ in the unstrained ML WSe$_2$ shows that the strongest enhancement arises from the 2-p HX resonance state.
Subsequently, we strain-tune the excitonic absorption energy relationship to achieve  double resonance  ($E_{\rm{HX}}$ = 2$E_{\rm{A}}$).
A tensile biaxial strain of $0.16\%$ leads to a maximum ($50$-fold enhancement) in $\chi_{yyy}$ compared with the 2-p HX resonance state of the unstrained monolayer. 
Further exploration of the effect of strain reveals the existence of other double resonance states near the K valley that also enhance $\chi_{yyy}$.
Overall, our results suggest new possibilities for strain-tuning 2D material properties to greatly enhance optoelectronic properties for novel potential future optoelectronic applications.

$\it${Acknowledgment} - The authors acknowledge the financial support by the Research Grants Council of Hong Kong (YX, ZP, and DL- GRF Grant No. 15304519, DJS and ZG - GRF Grant No. 11211019).

\bibliography{reference.bib}
\end{document}